\documentclass[journal]{IEEEtran}
\pdfoutput=1
\usepackage{graphicx}
\usepackage[pdftex,plainpages=false,colorlinks,citecolor=black,linkcolor=black,urlcolor=black,pdfstartview=FitH,bookmarks=true]{hyperref}
\hypersetup{pdftitle={Spherical GEMs},pdfsubject={Firt results of Spherical GEMs},pdfauthor={Serge Duarte Pinto}}
\usepackage[varg]{txfonts}
\usepackage{upgreek}
\usepackage{graphicx}
\usepackage{bm}
\usepackage{footnote}
\usepackage{stfloats}
\usepackage{textcomp}
\usepackage[english]{babel}
\usepackage[tracking,spacing,kerning,letterspace=0,babel]{microtype}
\usepackage{wasysym}
\makeatletter
\renewcommand{\fnum@figure}{\figurename~\oldstylenums{\thefigure}}
\renewcommand{\thefootnote}{\oldstylenums{footnote}}
\renewcommand\@biblabel[1]{[\oldstylenums{#1}]}
\makeatother
\fussy
\usepackage{memhfixc}
\begin{document}
\title{First results of spherical \textsc{gem}s}
\author{\IEEEauthorblockN{Serge Duarte Pinto\IEEEauthorrefmark{1}\IEEEauthorrefmark{2}\IEEEauthorrefmark{3}, Matteo Alfonsi\IEEEauthorrefmark{1}, Ian Brock\IEEEauthorrefmark{2}, Gabriele Croci\IEEEauthorrefmark{1}, Eric David\IEEEauthorrefmark{1},\\  Rui de Oliveira\IEEEauthorrefmark{1}, Leszek Ropelewski\IEEEauthorrefmark{1}, Miranda van Stenis\IEEEauthorrefmark{1}, Hans Taureg\IEEEauthorrefmark{1}, Marco Villa\IEEEauthorrefmark{1}\IEEEauthorrefmark{2}.}
\thanks{Manuscript submitted to \textsc{ieee} Nuclear Science Symposium Conference Record, \oldstylenums{19} November \oldstylenums{2010}.}
\thanks{\IEEEauthorblockA{\IEEEauthorrefmark{1}\textsc{Cern}, Geneva, Switzerland.}}
\thanks{\IEEEauthorblockA{\IEEEauthorrefmark{2}Physikalisches Institut der Universit\"at Bonn, Bonn, Germany.}}
\thanks{\IEEEauthorblockA{\IEEEauthorrefmark{3}Corresponding author: Serge.Duarte.Pinto@cern.ch}}}

\maketitle
\thispagestyle{empty}
\renewcommand{\thefootnote}{\arabic{footnote}}
\noindent\begin{abstract}
\ We developed a method to make \textsc{gem} foils with a spherical geometry.
Tests of this procedure and with the resulting spherical \textsc{gem}s are presented.
Together with a spherical drift electrode, a spherical conversion gap can be formed.
This eliminates the \emph{parallax error} for detection of x-rays, neutrons or \textsc{uv} photons when a gaseous converter is used. 
This parallax error limits the spatial resolution at wide scattering angles.

Besides spherical \textsc{gem}s, we have developed curved spacers to maintain accurate spacing, and a conical field cage to prevent edge distortion of the radial drift field up to the limit of the angular acceptance of the detector.
With these components first tests are done in a setup with a spherical entrance window but a planar readout structure; results will be presented and discussed.

A flat readout structure poses difficulties, however.
Therefore we will show advanced plans to make a prototype of an entirely spherical double-\textsc{gem} detector, including a spherical \oldstylenums2\textsc{d} readout structure.
This detector will have a superior position resolution, also at wide angles, and a high rate capability.
\end{abstract}

\section{Introduction}
\begin{figure}
\center\includegraphics[width=.8\columnwidth]{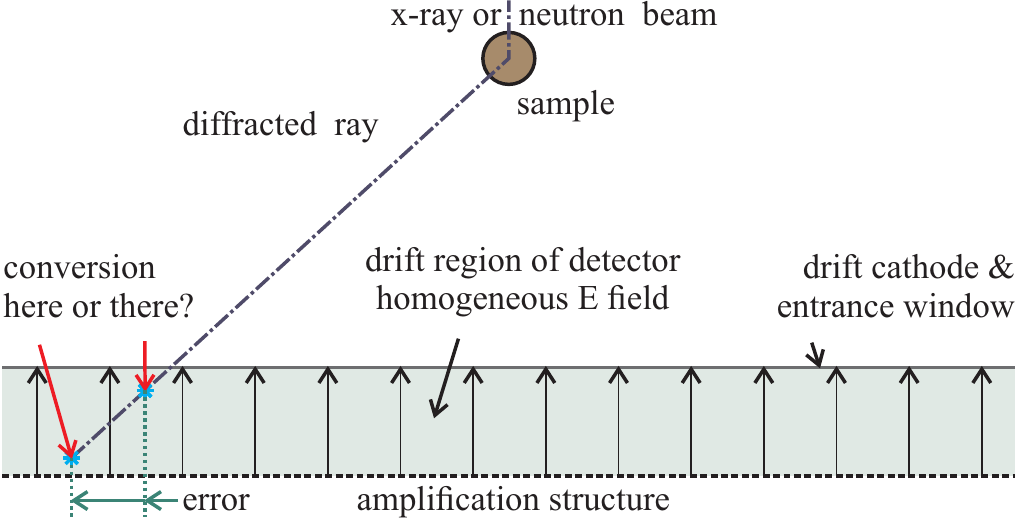}
\caption{The cause of a parallax error in a gas detector with a homogeneous drift field.}
\label{ParallaxError}
\end{figure}

\fnbelowfloat
\begin{figure}[b]
\includegraphics[width=\columnwidth]{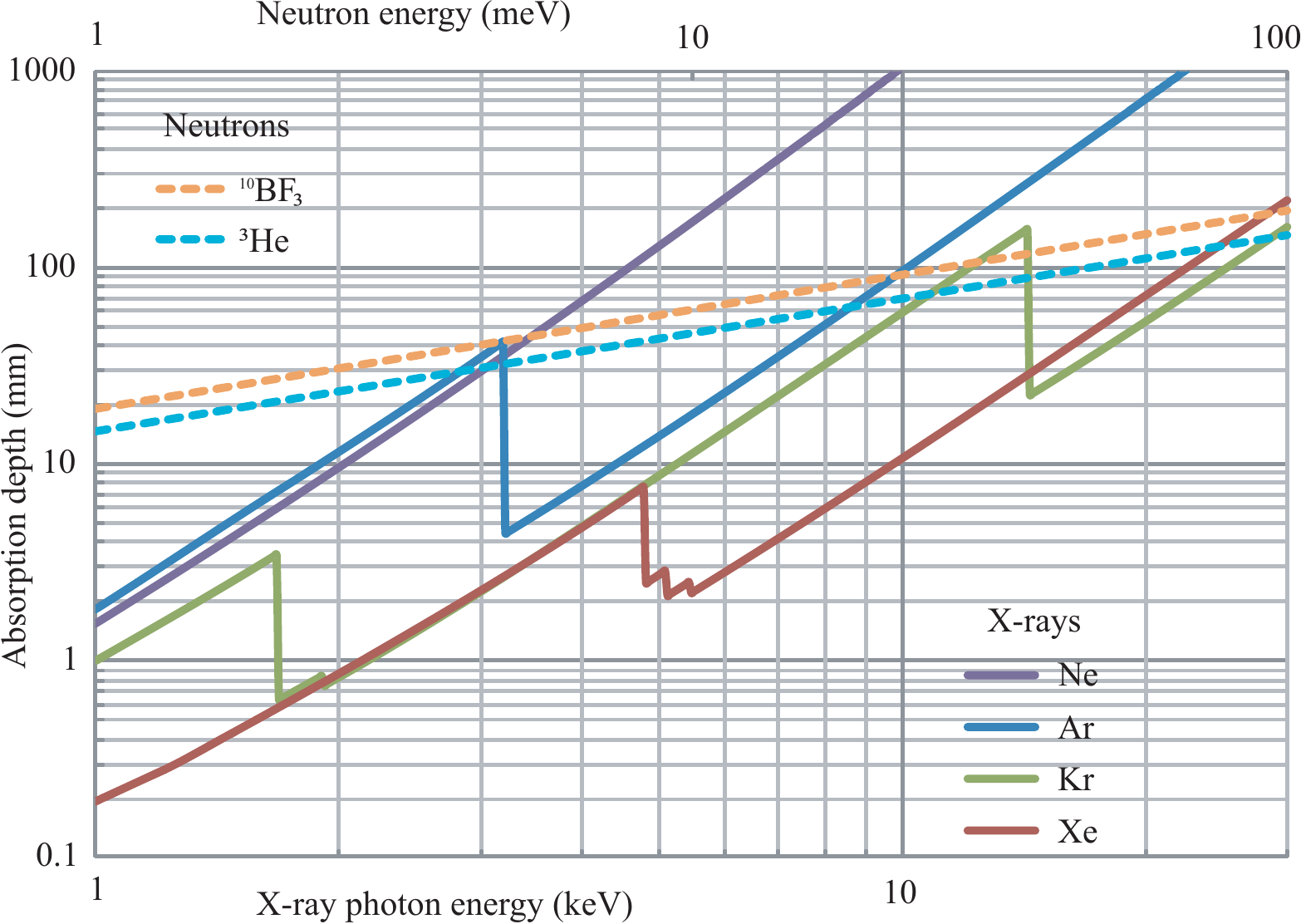}
\caption{Absorption length of various x-ray and thermal neutron conversion gases at atmospheric pressure, as a function of the energy of x-ray photons (lower horizontal scale) or neutrons (upper scale). Calculated from cross-section data~\cite{X-rayCross-sections, NeutronDataBooklet}.}
\label{Absorption_Gases}
\end{figure}
\label{intro}
\noindent The development of spherical \textsc{gem}s is aimed at eliminating the \emph{parallax error} in detection of neutral radiation (photons or neutrons) coming from a point-like source.
This error arises from the uncertainty of how deep radiation penetrates the sensitive volume before causing ionization (given that a gaseous component acts as a converter from neutral radiation to ionization).
Figure~\oldstylenums{\ref{Absorption_Gases}} shows that the depth of interaction of x-rays and thermal neutrons in a gas volume ranges from few to many millimeters, even when using the most efficient gases known for these purposes.
The situation is similar for gaseous \textsc{vuv} photoconverters such as tetrakis dimethylamine ethylene (\textsc{tmae}) or triethylamine (\textsc{tea})~\cite{CharpakParallax}.
If the electric field in the conversion region of a gas detector is not parallel to the direction of irradiation, an uncertain conversion depth will give rise to an error in position reconstruction, see Fig.~\oldstylenums{\ref{ParallaxError}}.

\begin{figure}[b]
\includegraphics[width=\columnwidth]{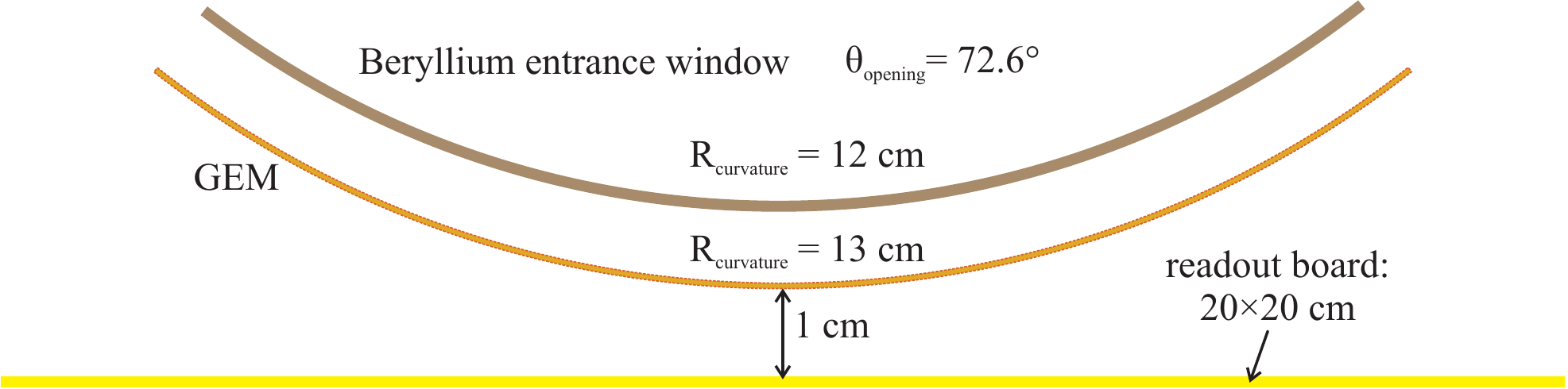}
\caption{Testing configuration of spherical \textsc{gem} in a chamber with a spherical beryllium entrance window and a flat readout.}
\label{SingleSpherical}
\end{figure}

Various methods have previously been used to suppress the parallax error \cite{LobsterPaper,CharpakPaper,Comparison,MoscowPaper,BrukerPaper,SegmentedCathodePaper,NeutronDiffraction}, each has its limitations.
In all cases the challenge of making a fully spherical detector, however desirable, is avoided.
With a spherical \textsc{gem} and a spherical cathode one can build a detector with a truly radial electric field in the conversion gap, thus eliminating the parallax error.
We developed a method to make spherical \textsc{gem}s, curved spacers and a conical field cage.
First tests of an assembly of these components were done in an existing detector with a spherical beryllium entrance window, see Fig.~\oldstylenums{\ref{SingleSpherical}}.

\section{Making spherical GEMs \textit{\&} other components}
\begin{figure*}
\center\includegraphics[width=.9\textwidth]{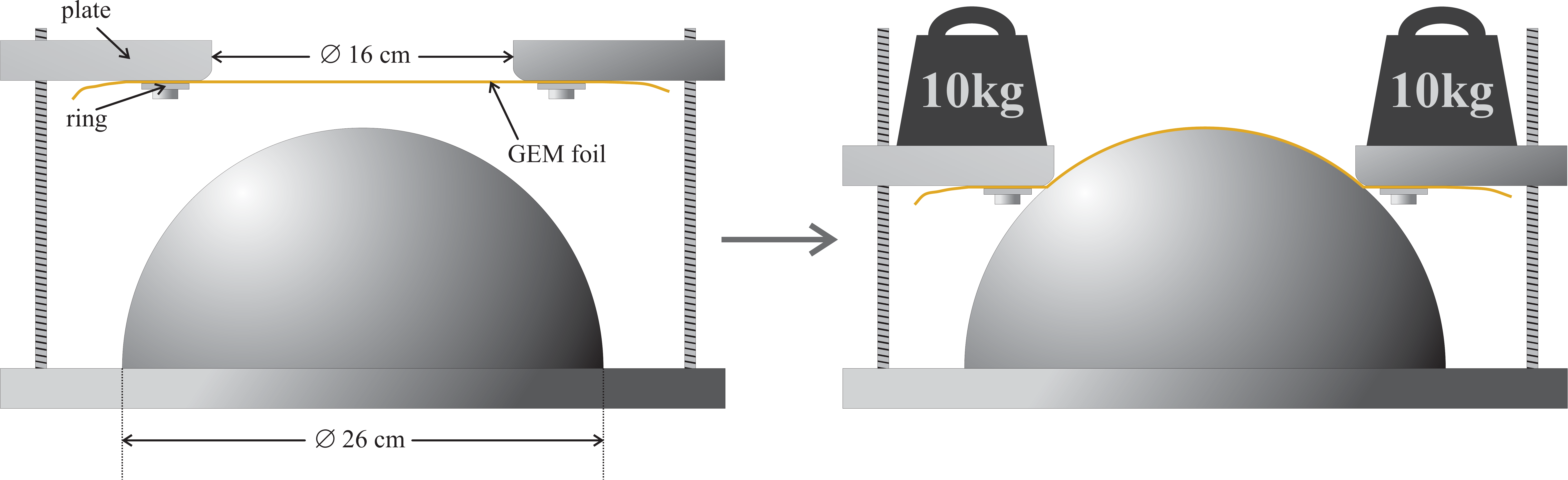}
\caption{The setup built to remold a flat \textsc{gem} into a spherical one. The setup is placed in a vacuum vessel inside an oven. All parts are stainless steel.}
\label{MoldingConstruction}
\end{figure*}
\noindent
A detailed description of the method with which spherical \textsc{gem}s are made can be found in~\cite{SphericalGEM_1,SphericalGEM_2}, here we only give an overview.
For the manufacturing of a spherical \textsc{gem} we start with a flat \textsc{gem} foil.
Major modifications to the manufacturing process are thus avoided.
Also, the \textsc{gem} foils can be tested before giving them their spherical shape, and afterwards the possible degradation of benchmark figures such as leakage current and discharge voltage can be checked.

Starting with this flat \textsc{gem} foil we use a method similar to thermoplastic heat forming; the foil is forced into a new shape by stretching it over a spherical mold, see Fig.~\oldstylenums{\ref{MoldingConstruction}}.
After a heat cycle it keeps this spherical shape.
Due to the thermoset nature of polyimide the forming process takes a long time ($\sim$\oldstylenums{24} hours) and requires a high temperature (\oldstylenums{350}--\oldstylenums{400}$^\circ$C).
In such conditions, the copper electrodes can oxidize completely, and delaminate from the polyimide.
Therefore, the whole forming procedure must be performed in a vacuum better than \oldstylenums{10}$^{-\oldstylenums3}$\, mbar.
A good vacuum also helps to prevent formation of deposits on the electrodes as a consequence of outgassing of polyimide and copper.

The spherical \textsc{gem}s made in this manner normally have the same discharge voltage (in air: $\ge$\oldstylenums{600}\,V) and leakage current ($\sim$\oldstylenums2\,nA) as before the heat forming treatment.
Surprisingly, \textsc{gem} holes observed through a microscope do not show any visible deformation, be it in the central area or closer to the edge.
This suggests that stresses are spread evenly across the foil and not concentrated in certain regions.
The increase in capacitance between the electrodes after changing shape is measured to be $\sim$\oldstylenums{11}\%, lower than predicted in \cite{SphericalGEM_2}: \oldstylenums{23}\%.
That prediction did not take into account that the holes are more strained by the stretching tension than the material between the holes.
The maximum opening angle we have been able to achieve with this setup is just over \oldstylenums{70}$^\circ$.

\subsection{Conical field cage}
\begin{figure}[b]
\includegraphics[width=\columnwidth]{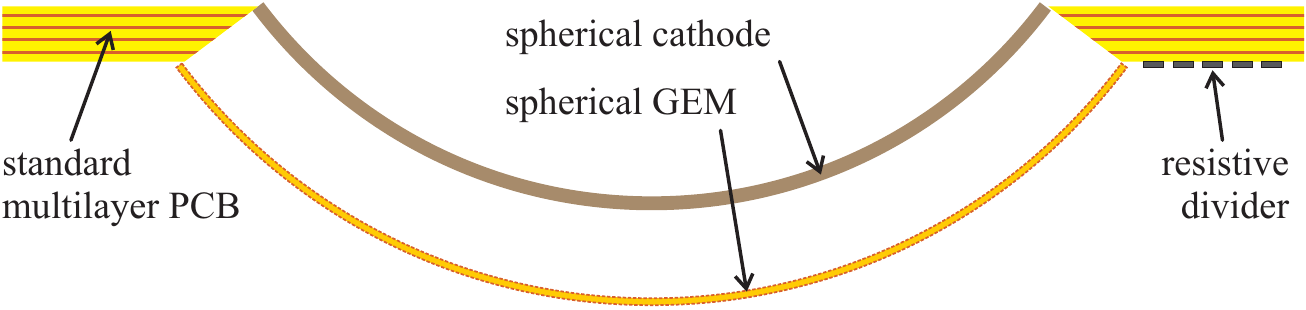}
\caption{The principle of a conical field cage made from a multilayer \textsc{pcb}.}
\label{FieldCage}
\end{figure}
\noindent
The field quality in the drift region is of critical importance, as the elimination of the parallax error depends on it.
We designed a field cage to maintain a good radial field until the edge of the active area.
This field cage is a conical enclosure of the conversion region, and is made from a standard multilayer \textsc{pcb}, see Fig.~\oldstylenums{\ref{FieldCage}}.
The inner metal layers serve as field electrodes.
A resistive divider defines the voltages supplied to each layer.
It also serves as a rigid mechanical fixture to which the \textsc{gem} can be glued, and as a high voltage distributor which supplies the \textsc{gem}.

\begin{figure}
\center\includegraphics[width=.9\columnwidth]{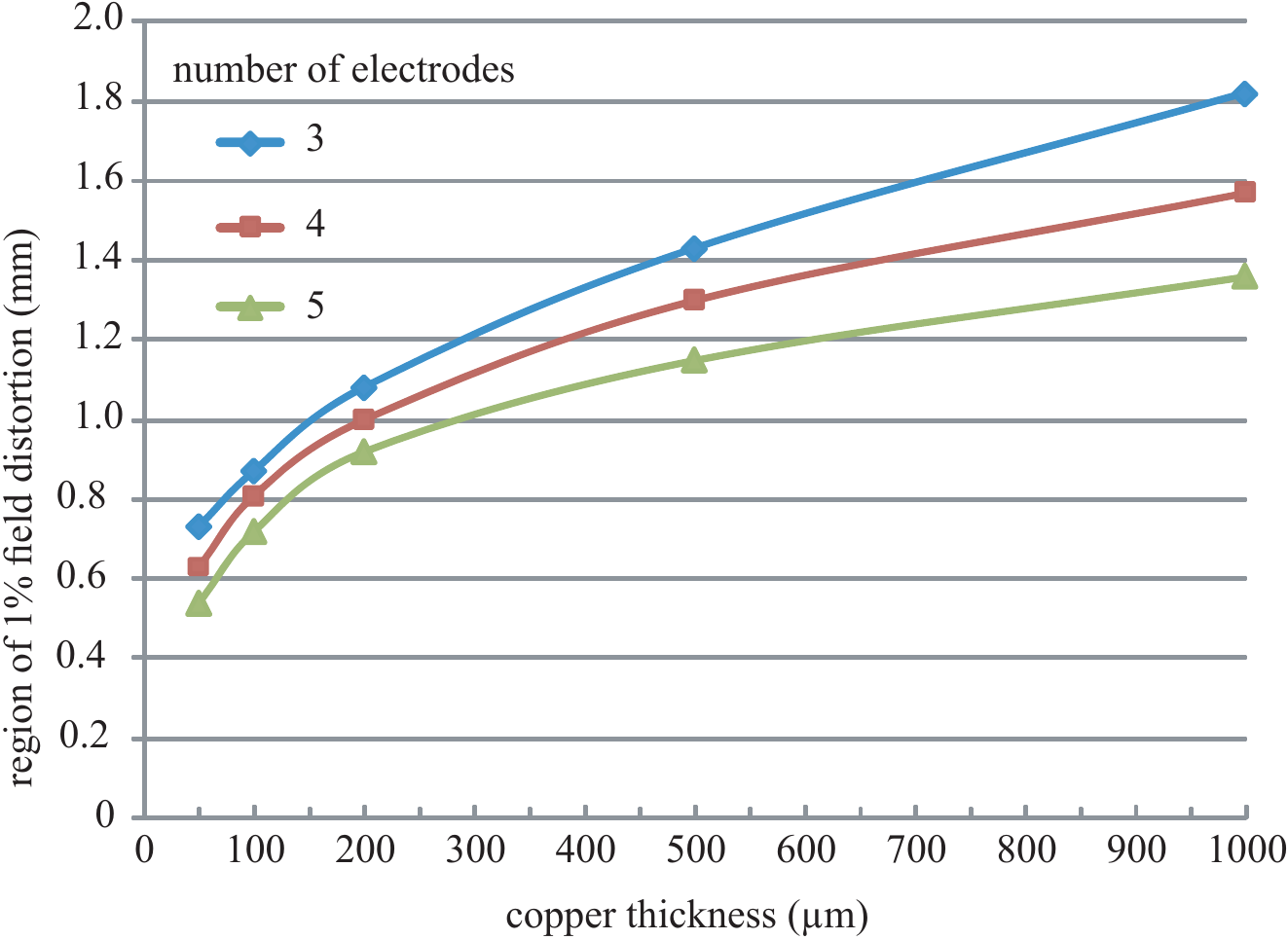}
\caption{Results of field simulations of a conical field cage. The width of the region with more than \oldstylenums1\% field distortion is plotted versus the thickness of the copper layers in the \textsc{pcb}, for the case of 3, 4 and 5 electrodes in the field cage.}
\label{FieldcageDistortion}
\end{figure}

Field simulations done with the Ansys\footnote{\href{http://www.ansys.com/}{http://www.ansys.com/}} fieldsolver indicate that the thin field electrodes (compared to the relatively wide strips commonly used in time projection chambers) do not give rise to field spikes close to the field cage.
Varying the thickness of the electrodes and the number of electrodes, the field distortions were simulated.
Figure~\oldstylenums{\ref{FieldcageDistortion}} shows that the fringes where field distortions are greater than \oldstylenums1\% of the nominal field steadily decline with increasing number of electrodes, and with decreasing electrode thickness.
The final prototype field cage we produced has 5 electrodes of $\sim$\oldstylenums{40}\,$\upmu$m thickness.

\subsection{Curved spacers}
\noindent
Although spherical \textsc{gem}s seem to be largely self-supporting, we developed curved spacers to add some rigidity and to tighten mechanical tolerances.
Flat spacers can easily be machined from plate material, but curved spacers are challenging both to design and manufacture.
We designed spacers to match our spherical \textsc{gem}s with Catia\footnote{\href
{http://www.3ds.com/products/catia}{http://www.\oldstylenums3ds.com/products/catia}} \oldstylenums3\textsc{d} design software.

The manufacturing was done using a technique called stereolithography, often used in industry for rapid prototyping of small, complexly shaped objects.
Objects are made out of a bath of liquid epoxy, which is selectively polymerized in a \oldstylenums2\textsc{d} pattern by a \textsc{uv} laser.
By consecutively making a large stack of such \oldstylenums2\textsc{d} patterns, the object takes shape.
The accuracy is often \oldstylenums{100}\,$\upmu$m or better in all three spatial coordinates.
There is a number of liquid epoxies suitable for this process, and one should choose the best in terms of rigidity, outgassing and dielectric strength.
\begin{figure}
\includegraphics[width=\columnwidth]{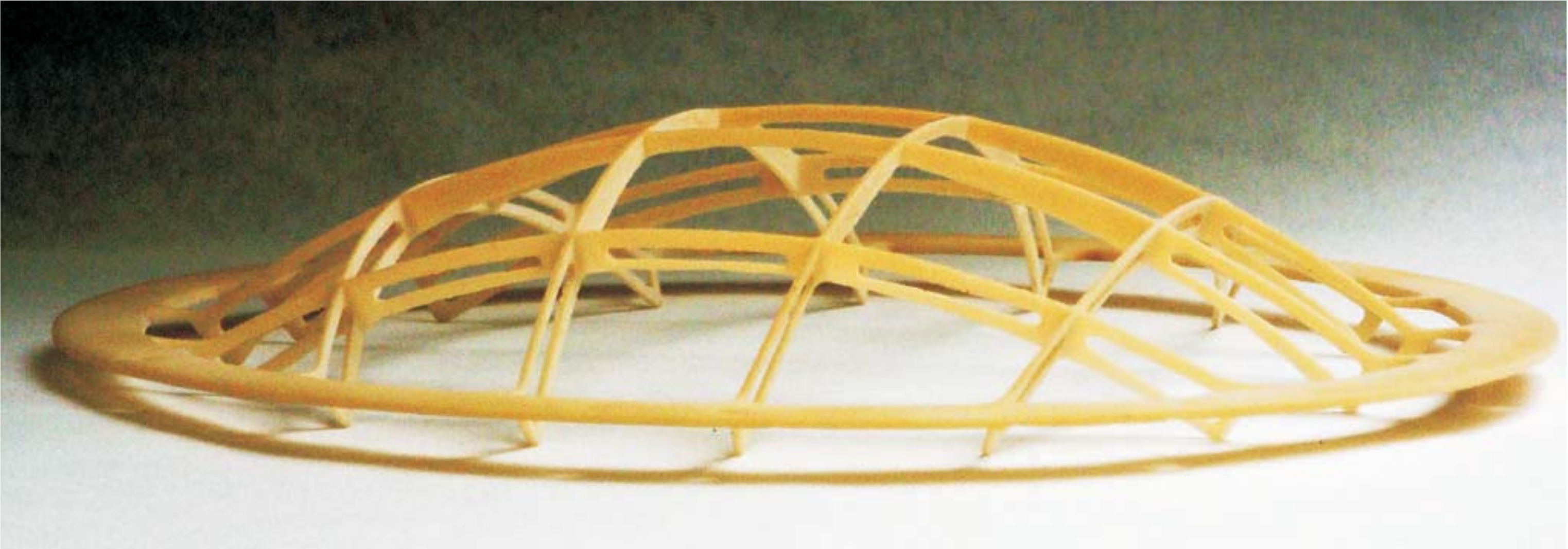}
\caption{A curved spacer made by stereolithography.}
\label{Spacer}
\end{figure}

\section{First results}
\begin{figure}[b]
\includegraphics[width=\columnwidth]{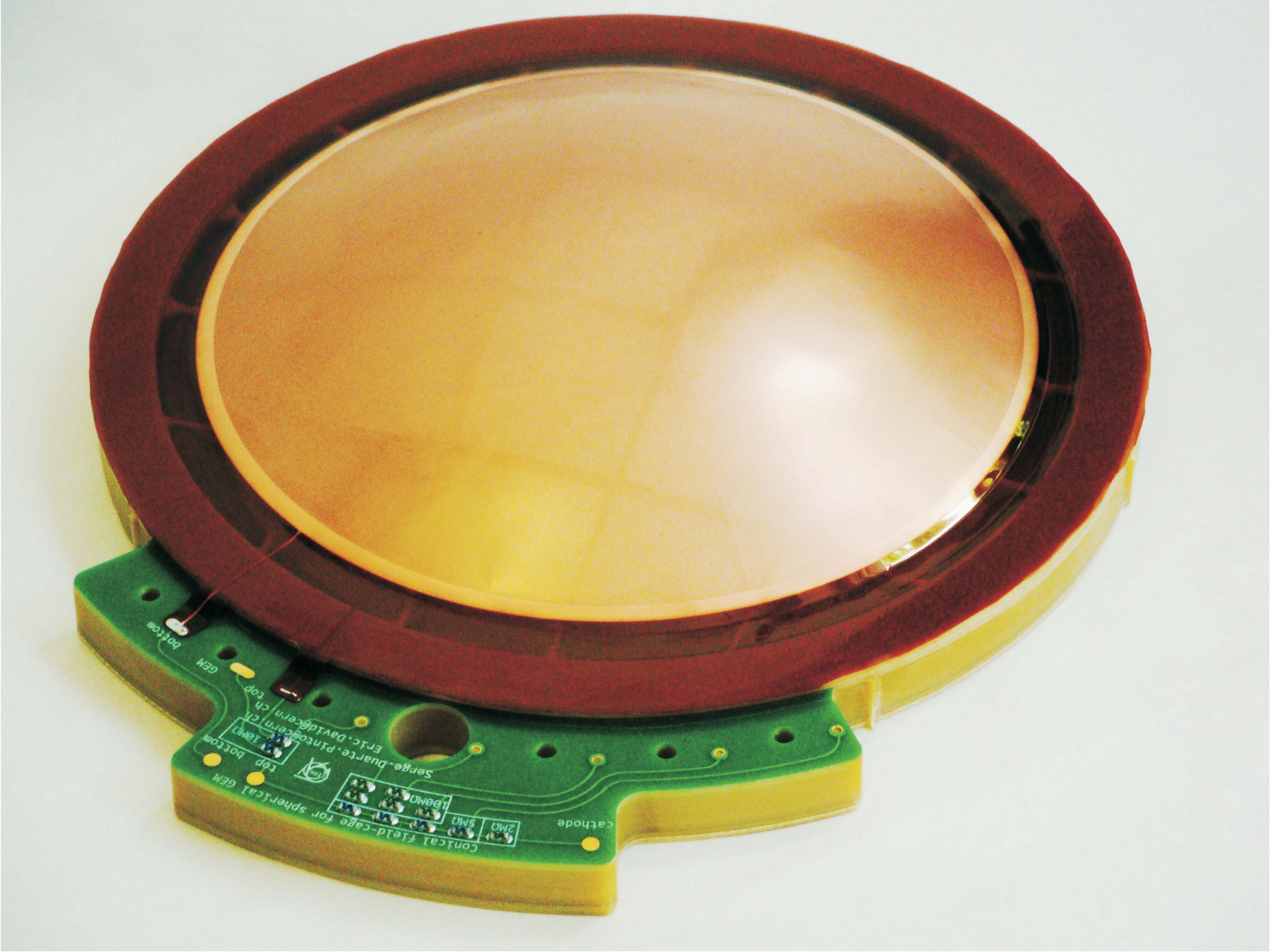}
\caption{An assembly of a spherical \textsc{gem}, a conical field cage and a curved spacer.}
\label{SphericalAssembly}
\end{figure}
\begin{figure}
\includegraphics[width=\columnwidth]{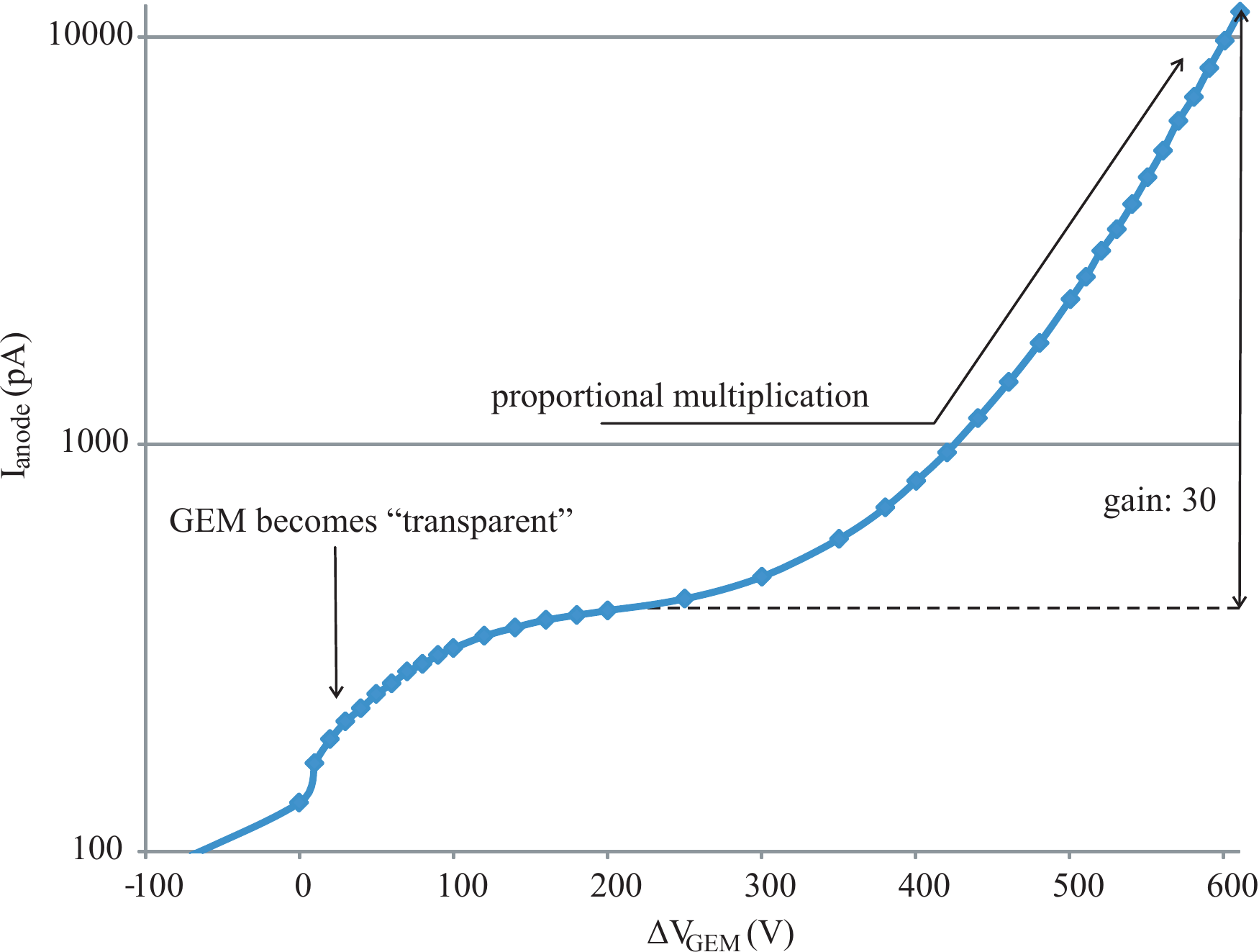}
\caption{Anode current of the setup shown in Fig.~\oldstylenums{\ref{SingleSpherical}} versus voltage applied to the \textsc{gem}. The gas mixture is Ar/CO$_{\oldstylenums2}$ \oldstylenums{70}\%/\oldstylenums{30}\% at an absolute pressure of \oldstylenums2 bar. Hardware issues with the detector in which the spherical \textsc{gem} assembly was installed kept us from raising the voltage further.}
\label{SphericalGain}
\end{figure}

\noindent
An assembly was made consisting of one spherical \textsc{gem}, a conical field cage and a curved spacer, see Fig.~\oldstylenums{\ref{SphericalAssembly}}.
This assembly was mounted in the detector mentioned above, with a spherical beryllium entrance window.
The detector needed some modifications in order to operate the \textsc{gem} in it, and some hardware issues are not entirely solved at the time of writing.
Therefore a full characterization of the properties has not yet been completed.
However, using a pA-meter typical \textsc{gem} behavior has been observed.
Figure~\oldstylenums{\ref{SphericalGain}} shows the anode current as a function of voltage applied to the \textsc{gem}.
At low voltage the \textsc{gem} becomes transparent to electrons coming from the conversion gap.
The anode current then reaches a plateau at full transparency, and when the voltage is further increased multiplication starts.
In the present conditions we report a gain up to \oldstylenums{30}; in general the spherical \textsc{gem} seems likely to perform just like a standard \textsc{gem}.

\section{Further development}
\begin{figure}[b]
\center\includegraphics[width=.8\columnwidth]{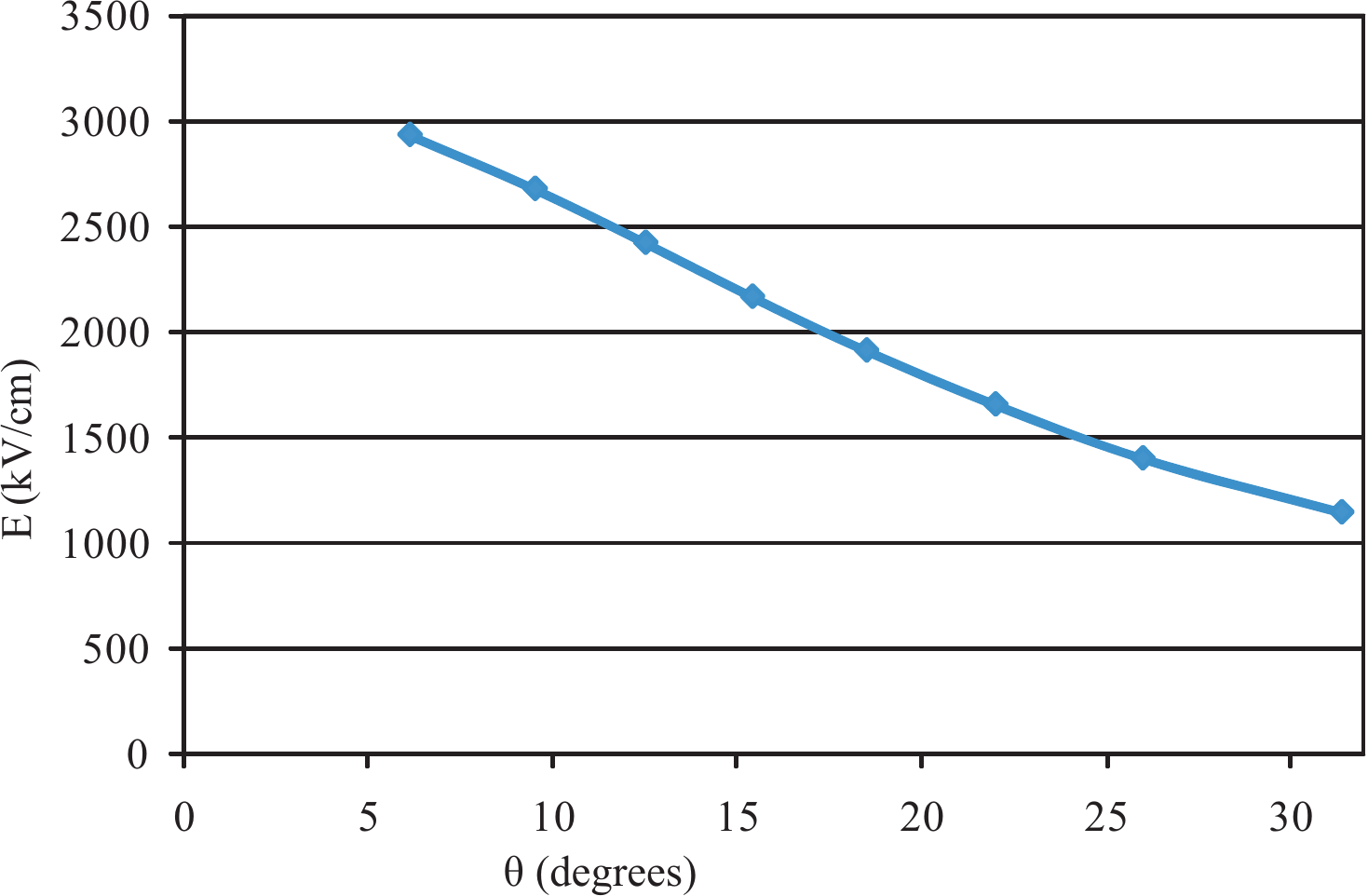}
\caption{Extraction field versus angle for the spherical \textsc{gem} with planar readout of Fig.~\oldstylenums{\ref{SingleSpherical}}.}
\label{ExtractionField}
\end{figure}
\noindent
Now that the feasibility of making and operating spherical \textsc{gem}s is demonstrated, we can start optimizing the readout structure.
In the present setup with a flat readout structure, the electric field between the spherical \textsc{gem} and the readout structure is rather nonuniform.
This is shown in Fig.~\oldstylenums{\ref{ExtractionField}}, where the field strength just below the \textsc{gem} in Fig.~\oldstylenums{\ref{SingleSpherical}} is plotted versus the angle with respect to the central axis.
This field extracts electrons from the \textsc{gem} holes, and influences the effective gain considerably.
\begin{figure*}[t]
\center\includegraphics[width=.89\textwidth]{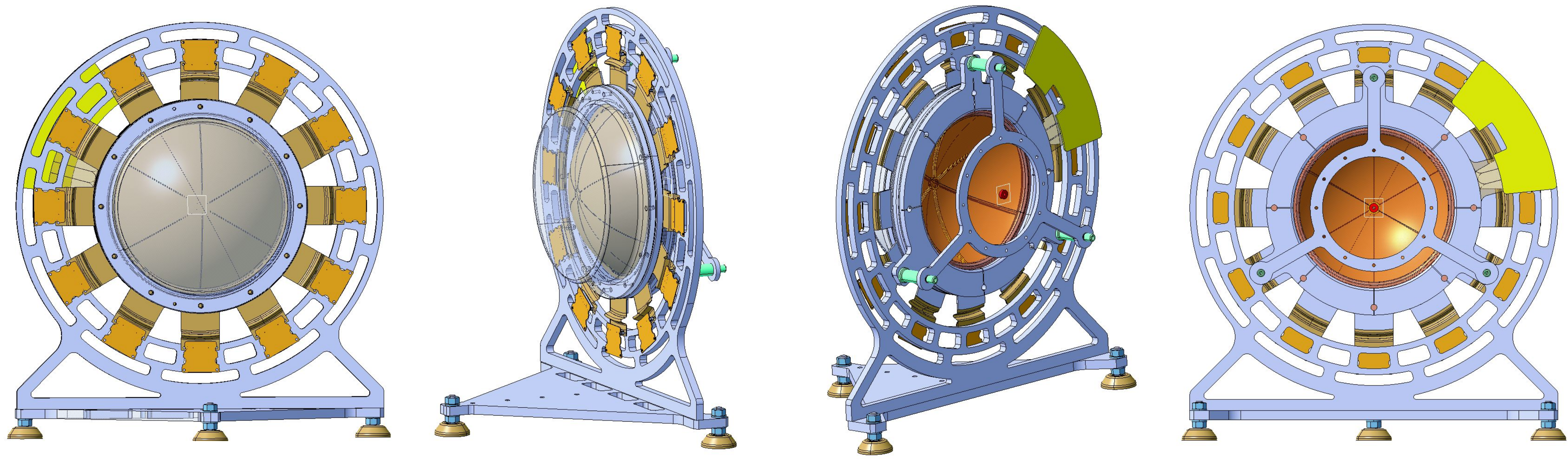}
\caption{3D design of a spherical multiple \textsc{gem} detector with spherical readout board from various points of view. The spherical active area has a diameter of $\sim 17$ cm, the whole structure is $\sim$ \oldstylenums{40} cm high.}
\label{3D}
\end{figure*}
Also, signal induction times vary considerably from the center of this detector to the periphery due to longer drift paths, and the weaker field in the periphery aggravates this effect.

The most obvious solution that comes to mind is making the readout structure spherical as well.
This is quite challenging, even now that we took the hurdle of making spherical \textsc{gem}s.
The most difficult step is the image transfer; no methods exist that allow transfer to spherical surfaces.
This was also the main reason to make spherical \textsc{gem}s from completely finished planar \textsc{gem}s.

Therefore we foresee making a planar \oldstylenums{2}\textsc{d} readout circuit from the same base material as \textsc{gem}s: copperclad polyimide.
Such a readout circuit we can form using the same tools and method as for spherical \textsc{gem}s.
There are still a few rather severe constraints, however.
Any readout structure with more than two metal layers requires some sort of adhesive or prepreg to make a mulitlayer board; no known adhesives or prepregs are compatible with the temperatures used while forming.
Even two metal layers can only be used if we can include vias in the structure; we have yet to prove that vias can survive the harsh heat forming treatment.

\begin{figure}[b]
\center\includegraphics[width=.9\columnwidth]{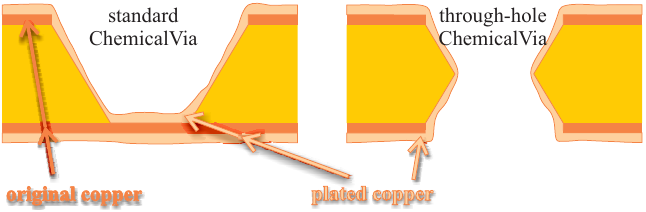}
\caption{Through-hole ChemicalVia (right) that is expected to survive a heat forming treatment, contrary to standard ChemicalVias (left).}
\label{Via}
\end{figure}

A microvia technology recently developed at \textsc{cern} (based on the \emph{ChemicalVia} technology developed earlier~\cite{ChemicalVia}) is expected to deliver vias that cannot be damaged by this treatment.
In a standard \emph{blind} ChemicalVia (see Fig.~\oldstylenums{\ref{Via}}, on the left) the adhesion of the plated copper layer to the copper that was originally part of the base material is not very strong, and could break when stretching and heating.
A \emph{through-hole} version of such a via can be made by plating a ``\textsc{gem}-style'' biconical hole; the contact between layers does not depend on the adhesion of the plated copper in this case.
A preliminary design of an \oldstylenums2\textsc{d} $r$-$\phi$ readout structure based on such vias (triplicated for redundancy) is made and ready for production.
Also mechanical designs for an all-spherical gas detector are in an advanced stage, see Fig.\oldstylenums~{\ref{3D}}.

\section{Conclusions \textit{\&} outlook}
\noindent
We developed spherical \textsc{gem}s, a conical field cage, and curved spacers, in order to make a parallax-free gas detector.
A first assembly is installed in an existing detector, which provides a spherical beryllium entrance window and a flat \oldstylenums2\textsc{d} readout board.
Familiar \textsc{gem} behaviour is demonstrated, consistent with standard flat \textsc{gem}s.
The observed gain of up to \oldstylenums{30} is expected to improve by one or two orders of magnitude when hardware issues with the detector are solved.

Plans have been made to develop also a spherical readout structure.
We made designs and identified technologies to make such a circuit, and also the mechanical structures necessary to build an all-spherical detector.

\section*{Acknowledgment}
\noindent
We would like to thank the team from the knowledge and technology transfer group at \textsc{cern} for their support for this project, as well as their successful negotiation of an agreement with an industrial partner.

\pagebreak
\end{document}